\documentclass[12pt]{article}
\usepackage[margin=1in]{geometry} 
\usepackage{amsmath,amsthm,amssymb,amsfonts}
\usepackage{graphicx}
\usepackage{hyperref}
\usepackage{cite}
\usepackage{xcolor}
\usepackage{ulem}
\usepackage{caption}
\usepackage{subcaption}
\usepackage{longtable}

\usepackage{setspace}

\usepackage{lineno}

\newcommand{\ttt}{\texttt}

\title{Sentinel node approach to monitoring online COVID-19 misinformation}
\author{Matthew T. Osborne, Samuel S. Malloy, \\ Erik C. Nisbet, Robert M. Bond, Joseph H. Tien}

\begin{document}
\maketitle

    \begin{abstract}
    	Understanding how different online communities engage with COVID-19 misinformation is critical for public health response, as misinformation confined to a small, isolated community of users poses a different public health risk than misinformation being consumed by a large population spanning many diverse communities.  Here we take a longitudinal approach that leverages tools from network science to study COVID-19 misinformation on Twitter.  Our approach provides a means to examine the breadth of misinformation engagement using modest data needs and computational resources.  We identify influential accounts from different Twitter communities discussing COVID-19, and follow these `sentinel nodes' longitudinally from July 2020 to January 2021.  We characterize  sentinel nodes in terms of a linked-media preference score, and use a standardized similarity score to examine alignment of tweets within and between communities.  We find that media preference is strongly correlated with the amount of misinformation propagated by sentinel nodes.  Engagement with sensationalist misinformation topics is largely confined to a cluster of sentinel nodes that includes influential conspiracy theorist accounts,  while misinformation relating to COVID-19 severity generated widespread engagement across multiple communities. Our findings indicate that misinformation downplaying COVID-19 severity is of particular concern for public health response.
    \end{abstract}

	\section{Introduction}\label{sec:intro}
		The proliferation of online misinformation has presented a challenge to public health throughout the COVID-19 pandemic \cite{brennen2020}, and is characterized by broad demographic and geographic reach \cite{gallotti2020}. Misinformation exposure reduces adherence to non-pharmaceutical interventions (NPIs) \cite{bridgman2020} and consequently has driven negative health outcomes in groups with high exposure to such content \cite{ash2020}.  The U.S. Surgeon General issued an advisory declaring health misinformation a ``serious threat to public health'' \cite{surgeon-general2021}. The significant potential for misinformation to drive behavioral change such as vaccine hesitancy \cite{loomba2021,bubar2021} underscores the importance of identifying and mitigating COVID-19 misinformation.
		
		Assessing the public health risk posed by specific types of misinformation is challenging in the low signal-to-noise environment of social media platforms, which are key mechanisms for COVID-19 misinformation spread \cite{allcott2017,evanega2020}. Since March 2020 tens of millions of tweets regarding COVID-19 have been posted daily \cite{lamsal2020,qazi2020}, of which only a fraction contain misinformation. Developing techniques for researchers and public health officials to identify misinformation that is circulating online as well as to distinguish which types of misinformation should be addressed quickly is of critical importance.
		
		Large-scale repositories of COVID-19 related tweets provide important resources  for identifying COVID-19 misinformation \cite{chen2020,deverna2021,gallotti2020,muric2021}.  The computational demands of working with these large datasets are, however, considerable, and determining which tweets correspond to misinformation is not straight forward. Moreover, a seemingly large volume of false or misleading COVID-19 content does not necessarily imply broad engagement. The breadth and depth of COVID-19 misinformation are important considerations: misinformation 
		confined to a small, isolated group of users has different public health ramifications than misinformation being consumed by a large portion of the population across different communities and demographic groups.  
		
		Here, we describe a longitudinal, network-based approach to characterize the breadth of online engagement with specific misinformation topics.  We focus on Twitter, an important platform for the dissemination of content and influencing opinion, including COVID-19 misinformation \cite{cinelli2020,gallotti2020,yang2021}. Twitter has a natural network structure, for example through retweets, followership and co-linkage to other domains. We utilize this network structure to detect so-called `communities', tightly-knit groupings of users and accounts, discussing COVID-19 on Twitter. Communities play a particularly important role in the propagation of online content. Content sharing within communities is a key mechanism for exposure to and amplification of viewpoints \cite{cinelli2021,flaxman2016}, and provides important context for breadth of misinformation penetration. Incorporating community structure and other network features into misinformation monitoring efforts can thus enhance the detection of false or misleading content that has the potential to become broadly disseminated and to drive behavior change. 
		
		From each community we selected and monitored influential accounts, referred to as `sentinels', from July 2020 to January 2021. Sentinel communities were characterized according to the link sharing behavior of their constituent accounts. Twitter community structure has been observed to be highly assortative with respect to media preferences \cite{tien2020}, which have been shown to be correlated with attitudes regarding COVID-19 \cite{jamieson2020, motta2020}. Characterization of communities in terms of media preference allows for the establishment of baseline metrics for the propensity to engage with and propagate COVID-19 misinformation, and comparison of tweet similarity across the media preference spectrum.  This is critical for assessing breadth of misinformation penetration, including identifying consequential events where specific false or misleading content propagates from a fringe group that regularly engages in conspiracy theories to a more mainstream community. 
		
		This structured `sentinel node' approach provides a way to compare tweets across segments of the Twitter ecosystem and identify misinformation topics with broad penetration, while only requiring modest data storage and computational resources. Further, our longitudinal approach helps address issues with selection bias and temporal changes in phrase and hashtag usage that can be problematic for cross-sectional studies \cite{tufekci2014}.
		
		We find evidence that the linked-media preferences of the sentinel accounts is strongly correlated with their propensity to post COVID-19 misinformation.  Accounts linking predominantly to right-leaning domains tended to post more COVID-19 misinformation than those linking predominantly to left-leaning domains.  
Importantly, we observe that sensational topics were largely confined to a subset of conspiracy minded communities, while misinformation on COVID-19 severity was much more widespread. This includes a large `virality' event that appears to be catalyzed by former President Donald Trump. These results indicate that perceived COVID-19 severity is of particular concern for public health.

	\section{Results}\label{sec:results}

		\subsection{Twitter communities and sentinel node identification}

			A query for tweets containing the term `covid' yielded 168,950 tweets (74\% retweets) sent on May 27, 2020 by 142,331 unique accounts (Methods, Section \ref{sec:meth-recruit}).  We constructed a weighted, directed graph with adjacency matrix $A$, where $A_{ij}$ is the number of times that account $j$ retweeted account $i$.   The in-degree of node $k$ is thus the number of times that $k$ was retweeted, and the out-degree of $k$ equals the number of times that $k$ retweeted other accounts.  We excluded self-loops (self-retweets).  The largest connected component of this `retweet network', denoted $G$, contained 78,680 nodes and 87,030 total retweets. 

			Modularity maximization using a Louvain method \cite{blondel2008}  yielded 148 communities  in $G$, ranging in size from 5,641 to 6 nodes (Methods, Section \ref{sec:meth-recruit}).   Let $C$ denote the largest 28 communities comprised primarily of English-speaking, domestic accounts.  Each community in $C$ contained at least 429 nodes,  from which we selected the 15 most-frequently retweeted accounts as {\it sentinel nodes} to follow longitudinally.  Note that the in-degree distribution for $G$ has a heavy tail: the 15 most highly retweeted nodes in a community account for on average 84\% of the total number of times that nodes in that community were retweeted (median 86\%; interquartile range 77\% to 96\%). We will refer to the set of sentinel nodes as $S$, and the sentinel nodes drawn from community $c$ in $G$ as $S_c$.

			To examine robustness of the community assignments, we constructed a second retweet network $\tilde{G}$ based upon the same search term over a second, later time interval (June 8-9, 2020).  All of the 28 communities in $S$ were represented in $\tilde{G}$ (in the sense that at least one node from each community in $S$ was present in $\tilde{G}$).  The Rand index comparing the Louvain-detected communities for nodes common to both $G$ and $\tilde{G}$ was 0.96 (z-score 897.7) \cite{traud2011}, indicating significant correlations in the community structure of $G$ and $\tilde{G}$ and consistent with robustness in the community structure of the Twitter conversation regarding COVID-19 over this time period. 

			The most recent 3200 tweets posted by each sentinel node in $S$ were collected at least once per week between July 1, 2020 and January 6, 2021, yielding a dataset of all $4{,}130{,}909$ tweets posted by the sentinel nodes over this time period, with the possible exception of deleted tweets.

		\subsection{Sentinel community characterization using linked domains}

			We characterized sentinel communities through the domains shared in their tweets posted from 7/1/2020 - 10/3/2020. We used principal components analysis (PCA) to derive a linked domain score for each of our sentinel communities. Notably, we made no assumptions on the bias or reliability of the shared domains, so any correlation between the linked media score with COVID-19 misinformation is not a contrivance of the dimension reduction process.
			
			\begin{figure}[h!]
			\centering
			\includegraphics[scale=.4]{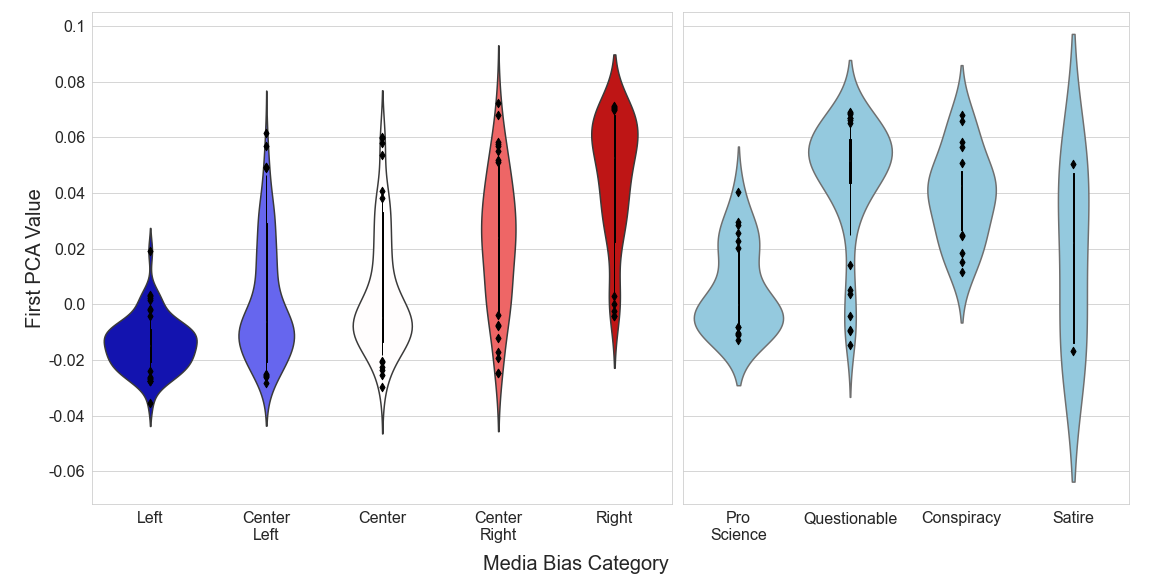}
			\caption{Distribution of linked domain first principal component entries by Media Bias Fact Chart (MBFC) categories \cite{mbfc}.  
			$38\%$ of linked domains from sentinel nodes were listed in MBFC as either Left, Center Left, Center, Center Right, or Right, and $9\%$ were listed as either Pro Science, Questionable, Conspiracy, or Satire. Treating the political categories as an ordinal variable with `Left' being equivalent to $1$, `Center Left'=$2$, `Center' = $3$, `Center Right' = $4$,  and `Right' $5$, there is a correlation of $0.66$ between the first PCA value and the political tilt of the domains represented.} 
			\label{fig:pca-violin}
			\end{figure}

			The $2{,}152{,}849$ tweets posted from 7/1/2020 - 10/3/2020 contained $706{,}564$ links. Of these, $147{,}510$ linked to Twitter and $44{,}529$ were shortened, yielding $514{,}525$ non-shortened links to $8{,}624$ distinct non-Twitter domains.   PCA on the domain frequency matrix (Methods, Section \ref{sec:meth-pca}) from each of the 28 communities in $C$  resulted in the distribution of domains along the first principal component value shown in Figure \ref{fig:pca-violin}.  Note that media sources with conservative-leaning partisan scores as described in \cite{mbfc} are associated with positive first component values (eg. \texttt{foxnews.com}, \texttt{oann.com}), while more liberal media sources are associated with negative entries (e.g. \texttt{motherjones.com}, \texttt{washingtonpost.com}). A table of the domains associated with the 30 most positive and negative first principal component values is given in the Supplementary Materials.  In the remainder we will refer to the first PCA score as the `Left' / `Right' linked domain score.

\begin{figure}[h!]
			\centering
			\includegraphics[scale=.25]{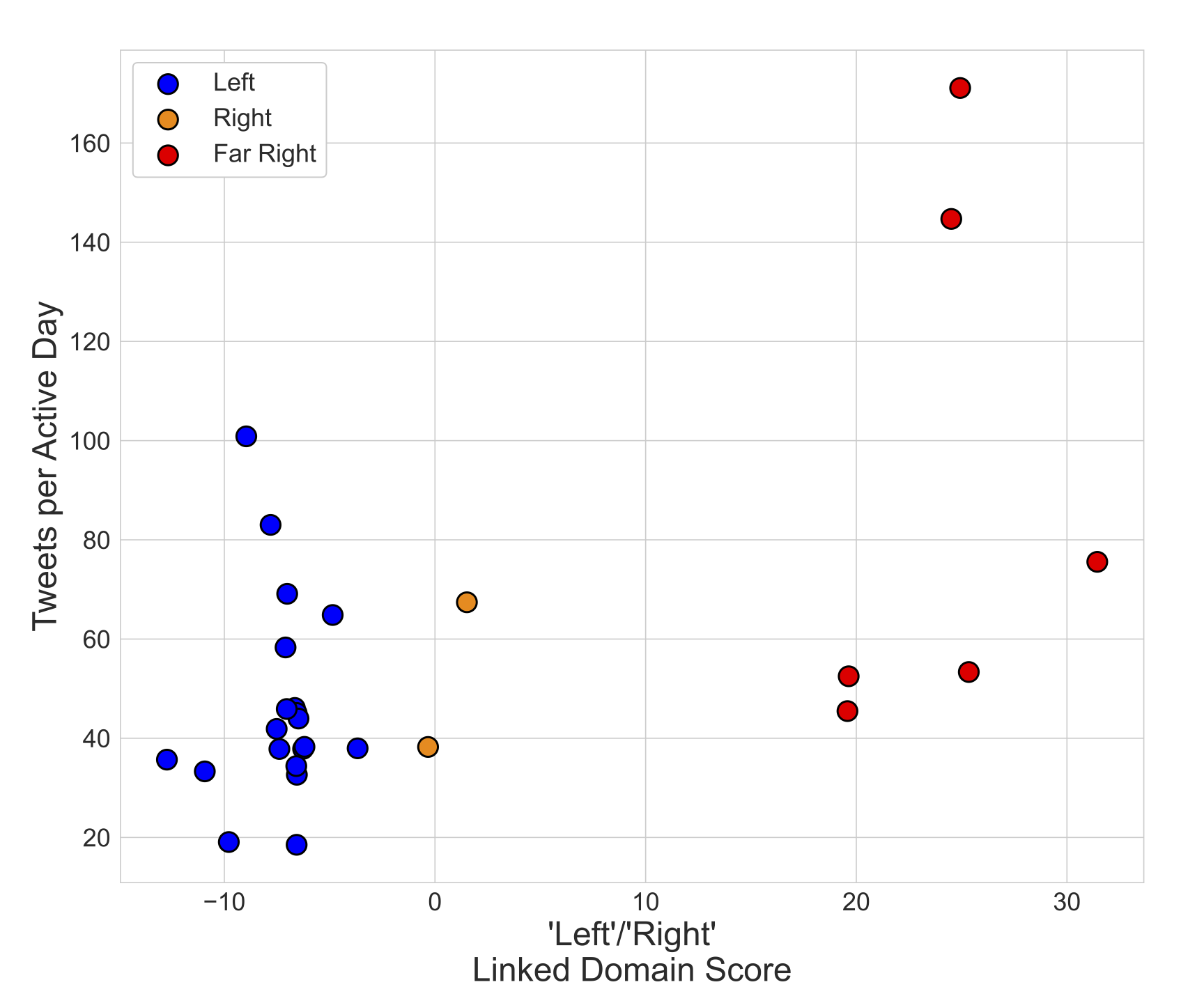}
			\caption{Sentinel communities projected onto the first component of the linked domain PCA space. Colors correspond to clusters resulting from mean linkage clustering on the first principal component score.  The vertical axis corresponds to the total tweets posted by the community over the observation period, normalized by the number of active days for the community's sentinel nodes (Methods, Section \ref{sec:meth-recruit})}
			\label{fig:pca}
			\end{figure}

			Left / Right linked domain scores for each sentinel node community $C$ are shown in Figure \ref{fig:pca}.  
			Hierarchical clustering (specifically, mean-linkage clustering \cite{mullner2013}) yielded three clusters of sentinel communities: a group with very negative Left / Right linked domain scores (blue nodes in Figure \ref{fig:pca}), a group with very positive scores (red nodes in Figure \ref{fig:pca}), and a group with intermediate scores (orange nodes, Figure \ref{fig:pca}).  The cluster with most negative (Left) linked domain score contains several Democratic politician accounts, while the cluster with most positive (Right) linked domain score contains some Republican politician accounts.  Profiles in the most positive cluster were, on average, about seven times more likely to post tweets containing strings associated with the QAnon conspiracy theory (e.g. ``qanon'', ``wwg1wga'', ``new world order'') than accounts in the other two clusters (see the Supplementary Materials).  We subsequently refer to the cluster with most positive linked domain score as `Far Right', and to the cluster with most negative linked domain score as `Left'. The cluster with intermediate linked domain score contains several pundits for mainstream conservative media.  Additionally, the linked domain score for this cluster was skewed left due to links to a domain connected with a specific sentinel node.  Removing such links increased this cluster's Left / Right linked domain score by about $10$ units without noticeably impacting the other two clusters. We thus refer to this cluster as `Right'.

		\subsection{COVID-19 misinformation by cluster}\label{sect:coding}

			To compare COVID-19 misinformation prevalence in tweets from the Left, Right and Far Right clusters, we randomly sampled 1,151 tweets stratified by cluster and topic.  The four topics examined were  COVID-19 mortality, masks, hydroxychloroquine, and Plandemic.  Tweet identification utilized a substring search (Methods, Section \ref{sec:meth-coding}).  
			
			Figure \ref{fig:pca-violin} indicates that the `Conspiracy' and `Questionable' media bias categories tended to be associated with more positive (Right-leaning) linked domain scores, suggesting that tweets containing misinformation may be more prevalent in the Right and Far Right clusters. To test whether prevalence differed between clusters, we developed a coding sheet with factual background information on the most common false or misleading claims associated with the aforementioned topics (see Supplemental Material).  Four human coders employed the reference sheet to individually annotate (1=misinformation, 0=no or unsure) whether each tweet presented misinformation.  Coding results yielded a Krippendorff's alpha of 0.73, indicating adequate inter-coder reliability \cite{krippendorff2013}.

			The summative results of the human coding are presented in Table \ref{table:coding}.  In total, 14.4\% of the tweets from the Left cluster on these four topics contained misinformation, compared to 85.1\% and 88.2\% of tweets from the  Right and Far Right clusters, respectively.  This difference between the three clusters was significant, $\chi^2(2, N=1151) = 563.3, \ p<.001$,  rejecting the null hypothesis of no difference in misinformation prevalence between clusters. Additional analysis indicated the difference in misinformation prevalence between the Right and Far Right clusters was not significant, $\chi^2(1, N=790) =1.7, \ p=n.s.$.  There was substantial variation in misinformation prevalence by topic, with misinformation appearing less frequently in tweets about face masks (54.1\%) and death severity (56.3\%) in all three clusters as compared to those about hydroxychloroquine (69.5\%) and Plandemic (79.6\%). 
			\begin{table}[ht]
			\begin{centering}
			\tiny
			\begin{tabular}{lcccccc}
			  & {\bf \# of} & {\bf Facemasks} & {\bf COVID-19 Mortality} & {\bf Hydroxychloroquine} & {\bf Plandemic} & {\bf Total} \\ 
			 {\bf Cluster} & {\bf Tweets} & {\bf \% Misinformation} & {\bf \% Misinformation} & {\bf \% Misinformation} & {\bf \% Misinformation} & {\bf \% Misinformation} \\ \hline
			Left & 361 & 9.5 & 14.0 & 19.0 & 16.1 & 14.4 \\
			Right & 382 & 72.0 & 77.0 & 95.0 & 98.8 & 85.1 \\
			Far Right & 408 & 82.4 & 77.5 & 94.1 & 99.0 & 88.2 \\
			Total & 1151 & 54.1 & 56.3 & 69.5 & 79.6 & 64.0 \\
			\hline
			\end{tabular}
			\caption{Frequency of COVID-19 misinformation within Left, Right, and Far Right clusters by topic.  For each topic (facemasks, COVID-19 mortality, hydroxychloroquine, Plandemic), the listed percentage denotes the proportion of the corresponding cluster's tweets on that topic containing misinformation. A small number of tweets contained misinformation regarding multiple topics.}
			\label{table:coding}
			\end{centering}
			\end{table}

		%
		\subsection{COVID-19 vaccines and disease severity content}\label{sec:res-content-analysis}
		As described in Section \ref{sect:coding}, there is high prevalence of misinformation regarding COVID-19 mortality specifically and COVID-19 severity more broadly in tweets posted by the Right and Far Right clusters of accounts.  Abundant online vaccine misinformation has also been documented \cite{muric2021}.  COVID-19 severity perceptions and vaccination decisions are related: perceived risks of infection versus vaccination factor into vaccination decisions \cite{bauch2004,kirzinger2021}, and polls indicate that perceived low risk of severe outcome is a potentially important rationale for vaccine hesitancy \cite{kirzinger2021}.  

		Tweets about vaccines and COVID-19 severity from sentinel nodes were obtained by filtering COVID-related tweets on associated keywords and phrases. Example keywords and phrases included `death rate', `fatality rate', and `confirmed cases' for COVID-19 severity, and `vaccine', `pfizer', `moderna', and `johnson and johnson' for vaccinations. These two topics were further subset with additional keyword and phrase filters to identify tweets concerning vaccine hesitancy, vaccine misinformation, and tweets that downplayed COVID-19 severity. Example keywords and phrases included `will not take' for vaccine hesitancy, `change dna' for vaccine misinformation, and `lower than flu'  and `mild' for downplaying severity. Complete lists are given in the Supplementary Materials.  
		
		The COVID-19 severity substring search returned $27{,}832$ tweets from the Left, $1{,}997$ from the Right and $7{,}377$ from the Far Right. Of those, $1{,}322$ Left, $372$ Right and $1{,}831$ Far Right tweets contained phrases associated with downplaying COVID-19 severity. The vaccination substring search returned $22{,}022$ tweets from the Left, $1{,}781$ from the Right and $4{,}298$ from the Far Right. Of those, $933$ Left, $165$ Right and $528$ Far Right tweets contained vaccine hesitancy phrases, while $312$ Left, $19$ Right and $401$ Far Right tweets contained vaccine misinformation related phrases. 

\begin{figure}[h!]
		\centering
		\includegraphics[scale=.18]{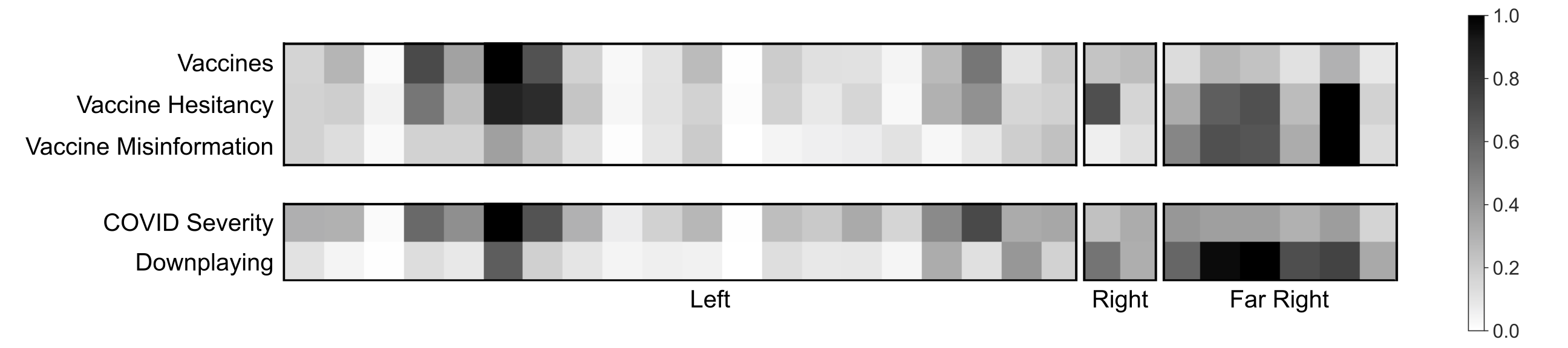}
		\caption{Distribution of topical tweets across communities.  Grayscale corresponds to the percentage of the maximum within subtopic per capita tweet rate  across communities. Per capita tweet rate corresponds to dividing the total number of topical tweets from each community by the number of active user days for the corresponding community.  `Vaccine Hesitancy' and `Vaccine Misinformation' tweets are subsets of `Vaccines' tweets; `Downplaying' tweets are a subset of `COVID Severity' tweets.  Phrases used for topic identification are given in  the Supplementary Materials.}
		\label{fig:topic-heatmap}
		\end{figure}

		Figure \ref{fig:topic-heatmap} compares tweet rates between communities for tweets concerning vaccines, vaccine hesitancy, vaccine misinformation, COVID-19 severity, and downplaying severity.  For each topic and community, we first compute the tweets per active account day for that cluster by dividing the total number of topical tweets by the number of active account days (Methods, Section \ref{sec:meth-recruit}).  We then scale this quantity for community $i$ by dividing the tweets per active account day for $i$ by the sum of the tweets per active account day across all communities.  This quantity is labeled the scaled per capita tweet rate for community $i$.  Each row of Figure \ref{fig:topic-heatmap} shows the scaled per capita tweet rates for a given topic.  

		We observed tweets containing vaccine hesitancy phrases across each cluster, though relatively few communities in the Left engaged with this topic.  Further examination of vaccine hesitancy tweets from the Left indicates that the majority of these tweets were commenting on `experimental' clinical trial progress for vaccine development, rather than expressing hesitancy to vaccinate.  Vaccine misinformation and severity downplaying tweets were largely confined to the Right and Far Right.  By contrast, scaled per capita rates for all vaccine and severity tweets were highest for the Left.  We thus observe selective higher engagement on vaccine hesitancy, vaccine misinformation, and downplaying severity from the Right and Far Right, despite lower engagement overall on vaccine and COVID-19 severity tweets compared to the Left.

\begin{figure}[h!]
			\centering
			\includegraphics[scale=.25]{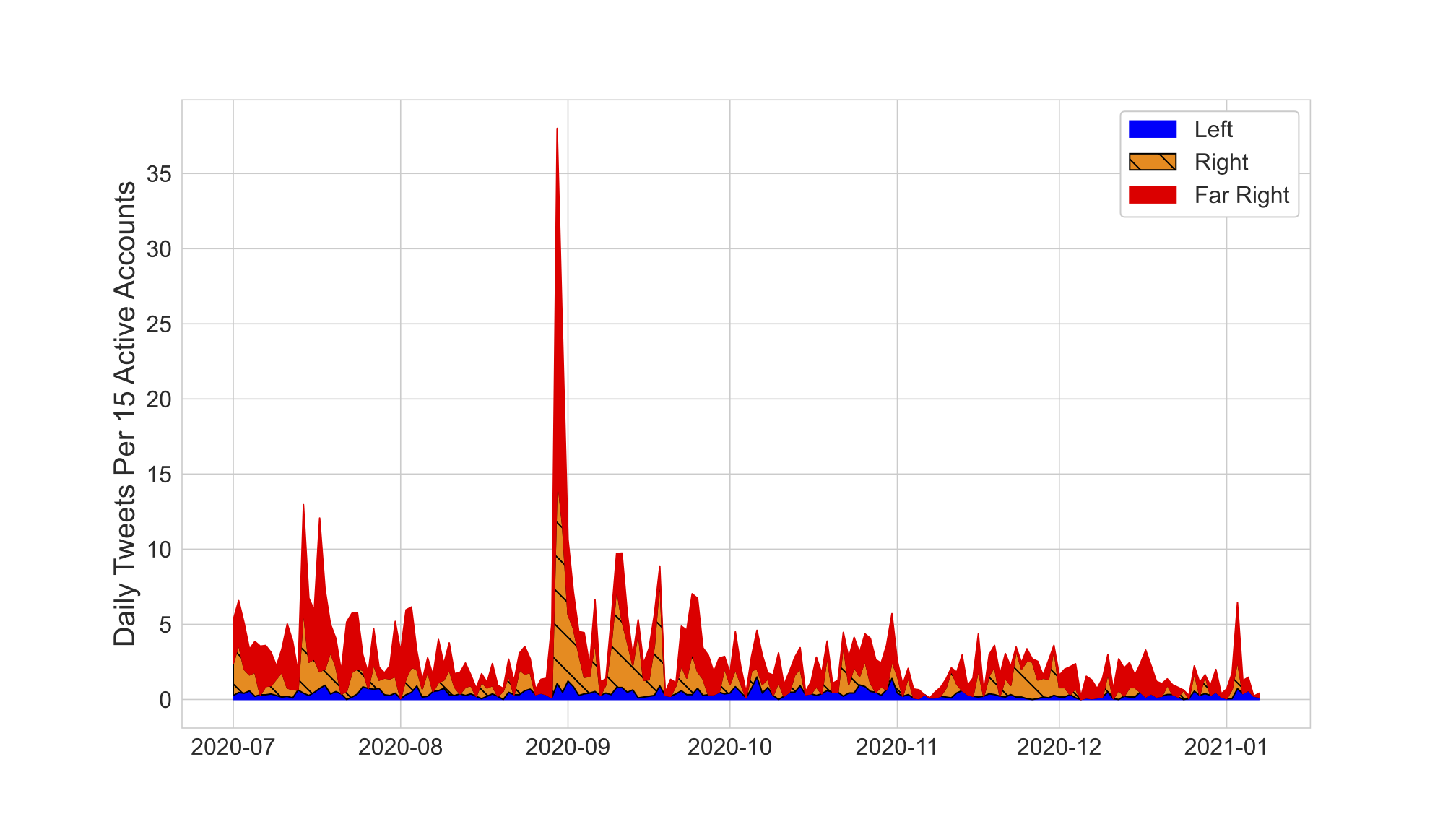}
			\caption{Daily tweets containing phrases associated with downplaying COVID-19 severity by cluster.  See  the Supplementary Materials for phrase lists.  Vertical axis corresponds to daily tweets per 15 active accounts.  Sustained high volume of tweets containing phrases downplaying COVID-19 severity are observed from the Far Right (red) and Right (orange) clusters of accounts. }
			\label{fig:severity-vs-t.png}
			\end{figure}

		\paragraph{Perceived COVID-19 severity.}
			A time series of daily tweets containing phrases downplaying COVID-19 severity aggregated by cluster is shown in Figure \ref{fig:severity-vs-t.png}.  We observe sustained high volume from the Far Right (red) and Right (orange) throughout the study period. We also observe occasional spikes, for example at the end of August.  These `bursts' are considered further in Section \ref{sec:res-viral}.

Widespread posting of tweets downplaying COVID-19 severity was observed across Right and Far Right communities, indicating cluster-wide engagement with the topic as opposed to engagement restricted to a handful of communities.  For all but one of the communities in the Right and Far Right clusters, more than 15\% of the tweets about COVID-19 severity contained phrases associated with downplaying COVID-19 risk.  The proportion of COVID-19 severity tweets that downplayed COVID-19 risk was over 70\% for one community in the Right.  

			A majority of downplaying tweets from the Right (59\%) and Far Right (75\%) contained phrases connected to the questioning of reported incidence and death statistics (Methods, Section \ref{sec:meth-content}). An inspection of these tweets reveals that both clusters tended to use words like ``overcount'', ``inflated'' and ``false positives''. They also highlighted various comorbidities identified in Centers for Disease Control (CDC) death statistics as the true causes of death. Further, a large fraction of tweets downplaying COVID-19 severity from the Right (47\%) and Far Right (30\%) contained phrases downplaying illness severity at the individual level. Such tweets tended to emphasize the survival rate for someone that contracts COVID-19. These results suggest that the perception that COVID-19 does not have a serious impact on the morbidity and mortality of the population or those individuals that contract the disease is not confined to a small subset of communities, but is widespread among the Right and Far Right clusters.

\begin{figure}[h!]
			\centering
			\includegraphics[scale=.21]{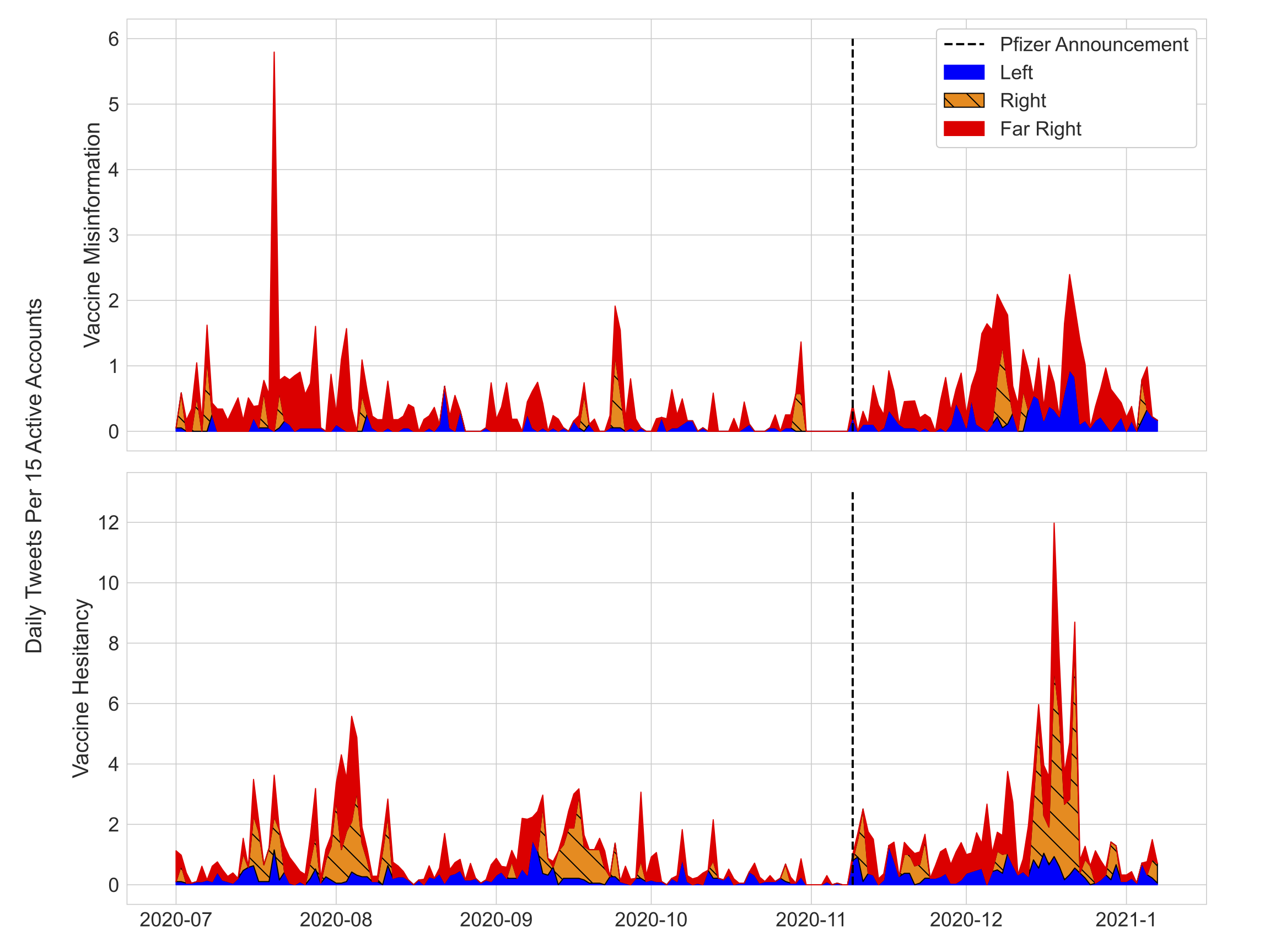}
			\caption{Daily tweets by cluster containing phrases associated with vaccine misinformation (Top) or vaccine hesitancy (Bottom).  Vertical axis corresponds to daily tweets per 15 active accounts.  Tweets containing vaccine misinformation phrases are primarily confined to the Far Right (red).  Tweets containing vaccine hesitancy phrases occur sporadically in both the Far Right (red) and the Right (orange), at a higher volume than in the Left (blue).  Vertical line corresponds to announcement of Pfizer phase three trial results on November 9, 2020.}
			\label{fig:vax}
			\end{figure}

		\paragraph{Vaccines.}
			Sentinel node tweets about COVID-19 vaccinations include both misinformation (e.g. ``the COVID vaccine will alter your DNA") as well as vaccine hesitant sentiments (e.g. ``the vaccine was rushed").  Both have the potential to undermine public vaccination campaigns.  Figure \ref{fig:vax} shows daily tweets by cluster containing vaccine misinformation phrases (Top) and vaccine hesitant phrases (Bottom) over the study period (Methods, Section \ref{sec:meth-content}).

			Tweets containing vaccine misinformation phrases were largely confined to the Far Right, which exhibited multiple days with high vaccine misinformation engagement relative to the other clusters. Three communities accounted for more than 70\% of all  Far Right vaccine misinformation tweets, suggesting heterogeneous engagement with such content. 
Inspection of these tweets indicates that vaccine misinformation content was not dominated by any one narrative.  Subtopics for vaccine misinformation tweets from the Far Right are given in Table \ref{table:far-right-cons-topics}. 
Apart from those tweets related to `Plandemic' (a conspiracy theory not solely confined to vaccines), no single topic accounts for greater than $15\%$ of the vaccine misinformation space. 			
			\begin{singlespace}
							\begin{table}[h!]
								\centering
								\begin{tabular}{l l}
									Vaccine Misinformation Topic & Percent of Misinformation Tweets \\
									\hline 
									Contains Microchips & \hfil$13.5\%$ \\
									Alters Your DNA & \hfil$7.3\%$ \\
									Will Sterilize You & \hfil$10.4\%$ \\
									Contains Aborted Cells & \hfil$11.8\%$ \\
									Depopulation/Genocidal Weapon & \hfil$14.7\%$ \\
									Plandemic & \hfil$32.11\%$ \\
									Other & \hfil$7.61\%$ \\ \hline
								\end{tabular}
								\caption{Prevalence of specific  subtopics in Far Right vaccine misinformation tweets. Percentages may surpass $100\%$ as a tweet can be flagged for more than one term or phrase.}
								\label{table:far-right-cons-topics}
							\end{table}
			\end{singlespace}
												
			Vaccine hesitancy tweets were more widespread among the Right and Far Right clusters compared with the Left (Figure \ref{fig:vax}, Bottom).  There are several days on which both these clusters engage with hesitancy content. Notably, these days occur both before and after Pfizer's November announcement of efficacy results from their phase 3 trial.  The percentage of vaccine-related tweets containing vaccine hesitancy phrases was comparable pre- and post-Pfizer announcement for the Right ($9\%$ pre, $10\%$ post) and Far Right ($14\%$ pre, $10\%$ post), but decreased by more than half for the Left ($7\%$ pre, $3\%$ post).  
			
			While the Right and Far Right both posted vaccine hesitant content, the flavor of content was slightly different between clusters. The Far Right's tweets were more likely to use phrases suggesting that the poster will not take the vaccine or urging others not to take the vaccine ($38\%$ of their total vaccine hesitant tweets compared to $6\%$ for the Right), while the Right was more likely to mention reports of adverse reactions ($42\%$ of their hesitant tweets compared to $13\%$ of the Far Right).

	\subsection{Flagging inter-cluster content spread}\label{sec:res-viral}
\begin{figure}[h!]
		\centering
		\includegraphics[scale=.2]{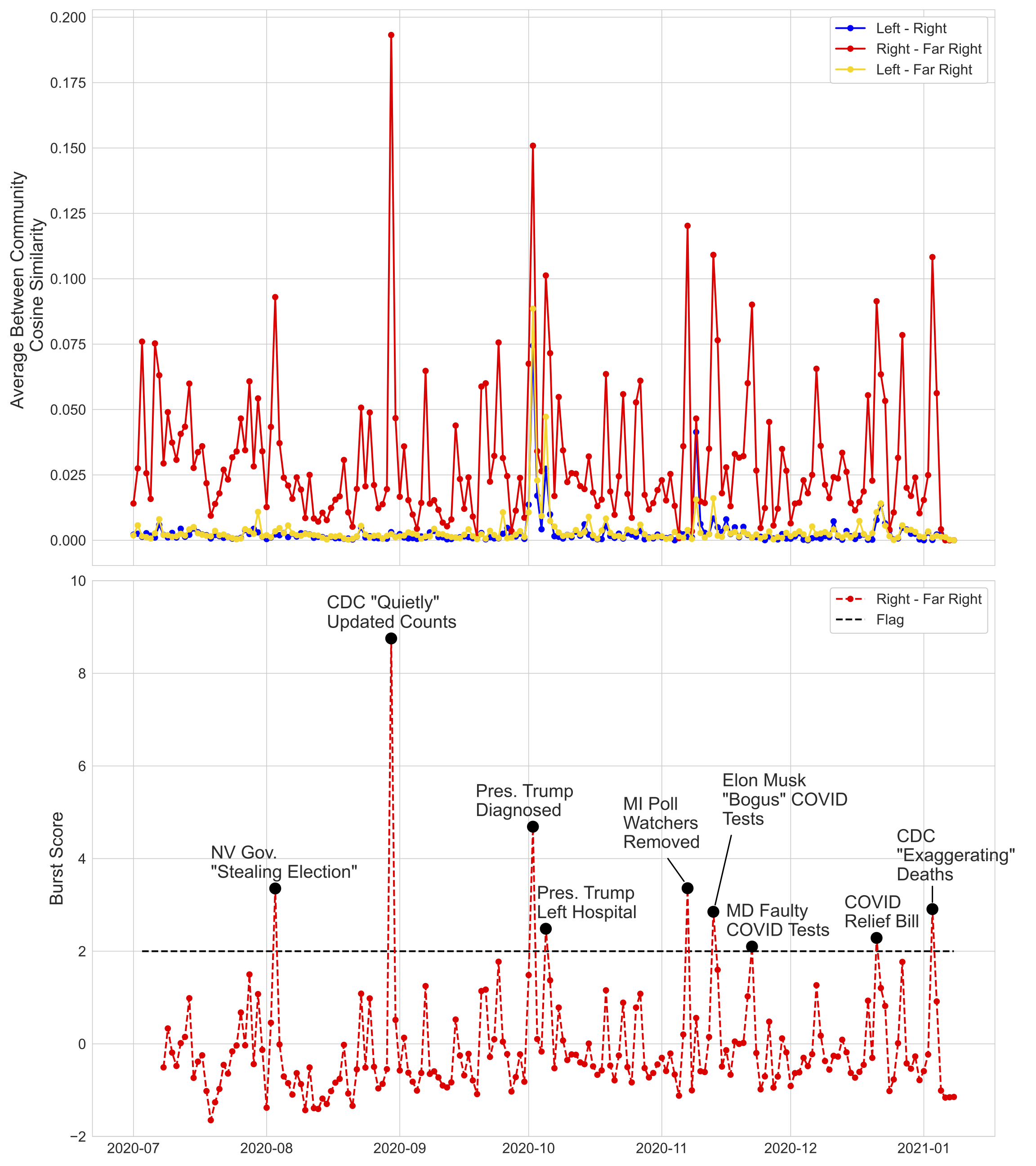}
		\caption{{\it Top.} Average inter-cluster similarity between the Left, Right, and Far-Right clusters for each day between July 1, 2020 and January 6, 2021.  {\it Bottom.} Inter-cluster burst score [Equation \eqref{eq:hist-sim}] between the Right and Far Right.  Annotated days represent days in which the burst score is greater than or equal to $2$.  The first seven days of the observation period are omitted to establish baseline. }
		\label{fig:sim}
		\end{figure}

		Figure \ref{fig:sim} (Top) shows trigram cosine similarity between pairs of clusters over time (Methods, Section \ref{sec:meth-sim}).  The inter-cluster similarity between the Left and the other two clusters is nearly zero for all but a handful of days.  By contrast, the Right and Far Right clusters have a non-zero baseline similarity and exhibit multiple days of similarity exceeding $0.10$.  Additionally, time points are observed where similarity spikes between the Right and Far Right.  These bursts may represent emerging content garnering support across different segments of the online ecosystem.

		We used a modification of the burst score of Mehrotra et al \cite{mehrotra2013} to identify time points for which inter-cluster similarity is unusually large.  Define 
		\begin{equation}\label{eq:hist-sim}
		H(A,B,t) = \frac{s_t(A,B) - \text{mean}_{\tau < t}\left\lbrace s_\tau(A,B) \right\rbrace}{\text{SD}_{\tau < t}\left\lbrace s_\tau(A,B) \right\rbrace} ,
		\end{equation}
		where $s_t(A,B)$ is the similarity between clusters $A$ and $B$ on day $t$ and $\text{SD}$ denotes the standard deviation.  The burst score $H$ gives a measure of inter-cluster similarity on a given day relative to historical similarity.  

		Figure \ref{fig:sim} (Bottom) shows the Right-Far Right similarity burst score \eqref{eq:hist-sim}, with days where $H>2$ (above the dashed line) flagged for further examination.   {\it Topical tweets} for the flagged days were identified using latent semantic analysis (LSA \cite{deerwester1990}; Methods, Section \ref{sec:meth-sim}).   Brief summaries of these `topics' are shown in Figure \ref{fig:sim} (Bottom).  Removal of topical tweets resulted in burst scores below the flagging threshold, consistent with the high burst scores on flagged days being driven by their spread. More in depth information including detailed examination of flagged days can be found in the Supplementary Materials.

		Topical tweets from flagged days show three themes: downplaying COVID-19 severity (e.g. doubting CDC reporting statistics and questioning the validity of COVID tests), tying of COVID-19 to positions of the Republican party from the 2020 presidential election (i.e., opposition to mail-in ballots and deriding politicians critical of Donald Trump), and news associated with Donald Trump's COVID-19 diagnosis \cite{ap_trump_diagnosis}  and subsequent departure from Walter Reed Medical Center \cite{ap_trump_leaves}.  The highest burst score corresponds to a topical tweet on August 30, 2020 related to CDC reporting of COVID-19 death statistics.  This is noteworthy, given the sustained posting of tweets downplaying COVID-19 severity by the Right and Far Right throughout the study period (Section \ref{sec:res-content-analysis}), and motivates further examination of content associated with the August 30, 2020 flagged day in Figure \ref{fig:sim} (Bottom).

\begin{figure}[h!]
			\centering
			\includegraphics[scale=.25]{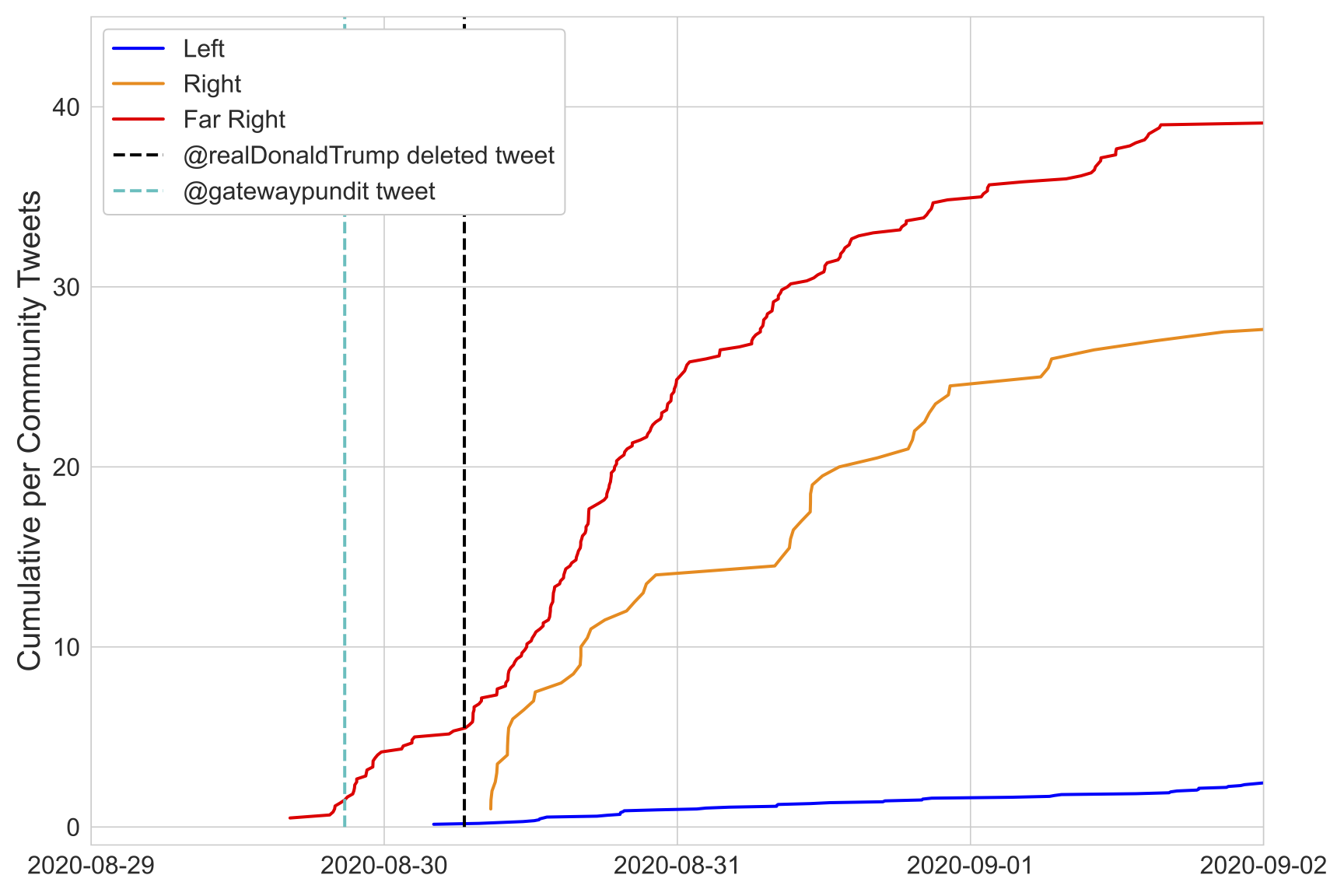}
			\caption{Cumulative per community tweet curves by cluster for tweets discussing the CDC altering its reported COVID-19 death statistics in August of 2020. The cyan dotted line denotes the time of a GatewayPundit tweet that sparked an amplification event in the Far Right and the black dotted line denotes the time of a deleted Donald Trump retweet of a QAnon account promoting the conspiracy.}
			\label{fig:cdc-lied}
			\end{figure}

		\paragraph{Example: inter-cluster spread event.}
			The high burst score on August 30, 2020 (Figure \ref{fig:sim}, Bottom) was driven by tweets claiming that the CDC had `quietly updated' COVID-19 death statistics indicating that only $6\%$ of the deaths previously categorized as being due to COVID-19 were actually caused by COVID-19, while the remaining $94\%$ were caused by underlying conditions \cite{ap_cdc_lied}.  Figure \ref{fig:cdc-lied} plots cumulative tweets on this topic from each cluster normalized by the number of communities in the cluster. Initial tweets occur in the Far Right, with early amplification within the Far Right corresponding to tweets linking to or retweeting the Twitter account of the Gateway Pundit (cyan dashed line).  Other studies have identified the prominence of the Gateway Pundit on Twitter generally and among fake news domains specifically \cite{lazer2020}. A subsequent amplification event occurs following a retweet by Donald Trump (black dashed line) of a known QAnon account on this topic \cite{ap_cdc_lied}.  The Right then begins to post on this topic. In this example, we thus observe a misinformation pathway that begins in the Far Right, includes accounts associated with conspiracy theorists, and subsequently spreads to more mainstream communities following amplification by an influential node.
				
	\section{Discussion}\label{sec:disc}
		The structured, data-light approach we have taken to monitoring COVID-19 misinformation is, to our knowledge, distinct from other published misinformation monitoring approaches to date.  Online communities within social media platforms vary widely in their posted content \cite{singh2020}, and previous work has shown that COVID-19 online content generally \cite{gallagher2021} and COVID-19 misinformation specifically \cite{yang2021} is driven by a small set of influential accounts. Identifying influential accounts from varied online communities to follow longitudinally leverages both of these facts.  The presented framework for structured, longitudinal monitoring can support early detection of narrative movement across communities, including noteworthy events such as narrative migration from a community with extremist tendencies into communities closer to the mainstream. This approach complements panel-based longitudinal studies that can be conducted at scale with regards to demographic information such as age, gender, and party affiliation \cite{lazer2020}. Principled selection of accounts translates into modest data requirements: we collected $4{,}130{,}909$ tweets posted by $420$ accounts over the six month period.  By contrast Chen et. al. \cite{chen2020}  pulled $764{,}613{,}007$ COVID-19 related tweets over the same time frame.

		The polarization of the U.S. electorate, the ramifications and connections with interactions and content on social media, and the politicization of the COVID-19 pandemic \cite{allcott2020,gollwitzer2020,green2020}, have been extensively documented.  Our finding that accounts sharing right-leaning media links more frequently posted COVID-19 misinformation compared with accounts sharing left-leaning links is consistent with other studies \cite{evanega2020,garrett2021,jamieson2020,lazer2020,motta2020}, and builds upon this existing literature to demonstrate a link between sharing right-leaning media on Twitter and sustained posting of COVID-19 misinformation. This is of particular concern given the influence of these sentinel accounts for COVID-19 discourse on Twitter. Polarization has been shown to impact compliance with recommended NPIs \cite{engle2020,grossman2020,painter2021}, and polling data suggests that it will be detrimental to mass vaccination efforts \cite{loomba2021,steelfisher2021}. 
		
		Translation of misinformation exposure to public health impact depends upon many factors, including extent of misinformation penetration, specific misinformation content, and demographic characteristics and social environment of the misinformation consumer \cite{druckman2021}.  Loomba et al \cite{loomba2021} demonstrate in an experimental setting that recent exposure to COVID-19 misinformation can produce statistically significant decreases among individuals' intent to vaccinate, although both the nature of the false or misleading content and the demographics of the information consumer can amplify or dampen this effect. The measurable impact of misinformation on vaccination rates is consequential given the continued emergence of SARS-CoV-2 variants of concern. Misinformation thus presents a critical threat in the persuasion of the vaccine hesitant, who play a key role in the direction and duration of the pandemic.
		
		Sensationalist narratives have been the focus of much discussion and do indeed present a potential risk to public health: among misinformation narratives presented to respondents, Loomba et al. find the claim that the COVID-19 vaccine would alter host DNA to be associated with the largest decrease in vaccination intent of the misinformation types considered \cite{loomba2021}. Hotez et al \cite{hotez2021} include `genetically modified humans' in their primer for healthcare providers for correcting COVID-19 vaccine misinformation.  We found that the genetic alteration, microchips, and Plandemic  COVID-19 conspiracy theories were largely confined to Far Right communities, and did not garner widespread cross-cluster support.  However, there was widespread penetration of misinformation downplaying COVID-19 severity, including cross-cluster propagation of content claiming manipulation of CDC death counts.  These latter findings are consistent with survey results \cite{druckman2021} and other Twitter studies \cite{jamison2020} which indicate widespread misconceptions of COVID-19 severity both online and offline.

		The prevalence of tweets downplaying COVID-19 severity across both the Right and Far Right clusters has public health implications. Romer and Jamieson \cite{romer2020} found that belief that the CDC exaggerated COVID-19 severity was associated with decreased vaccine willingness, and surveys released slightly after our observation period ended (February 2021) suggest that perceived COVID-19 severity is a key factor in an individual's decision to vaccinate \cite{nguyen2021,ruiz2021}.  Signaling from many thought leaders that COVID-19 is not a severe disease, despite reputable evidence to the contrary, could instill this position in Twitter users that consume our sentinels' content, especially if social media is their primary news source \cite{jamieson2020,bridgman2020}.  This may be particularly true if perceived vaccine risks outweigh the perceived risks from infection \cite{bauch2004}.
		
		Our sentinel approach is flexible, and can be extended to other topics and platforms (given data availability).  The basis of the method consists of identifying influential nodes in online communities, and examining content similarity over time between communities or clusters of communities.  We have intentionally taken a simple approach to measuring content similarity, using cosine similarity of trigrams.  This technique captures similarity driven by overlap in words (and thus, for example, captures retweet-driven similarity), but would not capture similarity of tweets using different words to express related content.  Using approaches such as word \cite{glove,word2vec,BERT} or tweet embeddings \cite{tweet2vec} to address this is an area for future work.

		Importantly, this analysis addresses the broader case of misinformation, which we consider to be false or misleading content regardless of intent, as opposed to the subset of misinformation known as disinformation which refers to intentionally disseminated false or misleading content within a target group to advance an agenda or to cause harm. The described framework could be used in conjunction with emerging techniques in the detection of influence operations, such as those developed by Smith et al. \cite{smith2021}, in order to explore the extent to which such actors drive meaningful narrative shifts across the social media ecosystem.

		A limitation of this study is its restriction to Twitter.  Each social media platform has its particular biases in user base and online functions, and study of consistencies and differences between platforms with regards to health information is important for misinformation monitoring and mitigation efforts.  Existing COVID-19 related work includes analysis of pro-vaccination and anti-vaccination content on Facebook \cite{sear2020}, popularity of YouTube COVID-19 misinformation and sharing of these videos via Facebook \cite{knuutila2020}, and cross-platform comparisons of Twitter and Facebook \cite{yang2021}.  Developing similar structured approaches to misinformation monitoring within and across additional platforms is an area for future work.
	\section{Methods}\label{sec:methods}
	
		\subsection{Sentinel recruitment and data collection}\label{sec:meth-recruit}
			\paragraph{Community structure and sentinel node identification.}	
				Queries for tweets containing the phrase `covid' were performed on May 27 over a twelve hour period using Twitter's API and the \ttt{tweepy} Python library.  Community detection was performed on the largest connected component of the retweet network, where the weight of the arc from $j$ to $i$ equals the number of times node $j$ retweeted $i$.  Specifically, we  maximize the following version of modularity for weighted, directed graphs: 
				\begin{equation}
				Q = \frac{1}{w} \sum_{i=1}^n \sum_{j=1}^n \biggl( A_{ij} - \frac{w_i^{in} w_j^{out}}{w} \biggr) \delta(C_i,C_j),
				\end{equation}
				where $w$ is the sum of all edge weights in the network, $w_k^{in}$ and $w_k^{out}$ the weighted in-degree and out-degree, respectively, of node $k$, $C_k$ the community assignment of node $k$, and $\delta$ corresponds to the Kronecker delta.  Modularity maximization was performed using a GenLouvain method \cite{blondel2008,genlouvain}, implemented with symmetrization of the modularity matrix $A_{ij} - \frac{w_i^{in} w_j^{out}}{w}$.  

				Sentinel nodes were selected by first considering the fifty largest communities in the retweet network, and then selecting the 15 most highly retweeted nodes from each of the communities that consisted of predominantly English-speaking, domestic accounts. Whether a community was predominantly English-speaking was determined by taking a random sample of $100$ tweets from that community and applying Google's language detection algorithm. Communities whose sample was at least 80\% English were classified as English-speaking. A final inspection was done on the filtered sentinel communities and any community whose sentinel accounts were clearly not based in the United States was removed.

			\paragraph{Stability of community structure.}
				To assess stability of community structure in Twitter content regarding COVID-19, we assembled a second retweet network using the same phrase search (`covid') over a 24 hour period beginning June 8 and ending June 9, 2020.  Modularity maximization was performed on this second retweet network, and then the  community structures of the May 27 and June 8 networks were compared using the Rand score as described in \cite{traud2011}.  Statistical significance of community similarity was assessed using the z-Rand score (\cite{traud2011}, equation 2.1).
		 
			\paragraph{Longitudinal data collection and observation period.}

				We collected tweets from sentinel nodes from July 1, 2020 through January 6, 2021 using Twitter's public-facing API.  This observation period was split into two sets using 10/4/2020 at 12:00 AM (ET) as the demarcation point. The first portion of the data (from 7/1/2020 - 10/3/2020) was used to characterize sentinel communities according to their linked domains (Section \ref{sec:meth-pca}) and establish baseline similarity between clusters (Equation \eqref{eq:hist-sim} and Section \ref{sec:meth-sim}).

			\paragraph{Node attrition, active accounts, and active user days.}
		    Node attrition over the observation period may occur due to several possible reasons, including a user deleting their account or suspension by Twitter.  Each cluster retained at least 80\% of their initial sentinel nodes through mid-December.  Comparison of attrition over time between clusters is given in the Supplementary Materials. In order to account for sentinel attrition we define the concept of an \textit{active account} as well as the corresponding notion of \textit{active account days}. We consider a sentinel account to be active on a given day if we observe a tweet from that account on or after that day. As an extension we define active account days to be the number of days a particular account is deemed active. 
		   		
		\subsection{Sentinel community characterization using linked domains}\label{sec:meth-pca}
			We examined the links that each community shared in their tweets posted from 7/1/2020 - 10/3/2020.  Specifically, we used principal components analysis (PCA) \cite{pca_paper} to produce a scalar measure of linked domain preference, and then clustered sentinel communities according to this preference. 
				
			A community-level domain frequency matrix was formed by examining the unique domains posted more than $10$ times by any given community. The $i,j$ entry of this matrix was the fraction of links from community $i$ that linked to domain $j$, excluding links to \texttt{twitter.com} and domains from URL shortening services (e.g. \texttt{bit.ly}). PCA was performed for this matrix and the output was cross-referenced with the media bias fact chart provided in \cite{mbfc}. 
				
			Sentinel communities were projected onto the resulting first principal component, giving each community a linked domain score.  Hierarchical clustering based upon the distance between cluster centroids \cite{mullner2013} was performed on the linked domain scores.  Cluster assignments were produced by selecting a cut point in the resulting dendrogram that yielded three clusters. To examine robustness of the identified clusters, we removed domains only linked to by a small fraction of users and reran the described clustering procedure.  Results of this robustness check are in the Supplementary Materials.
			
		\subsection{COVID-19 misinformation by cluster}\label{sec:meth-coding}

			COVID-19 misinformation by topic and cluster was determined through human coding of a random sample of tweets that were posted 7/1/2020-10/3/2020.  The considered topics were COVID-19 mortality, hydroxychloroquine, facemasks and Plandemic.

			Relevant tweets were selected with substring searches. First, a subset of tweets pertaining to COVID-19 was selected by finding any tweet that contained at least one of ``covid'', ``coronavirus'', ``sars-cov'' or ``pandemic''. These COVID-19 tweets were further subset for each of the four topics below by searching for the strings: ``plandemic'' and ``scamdemic'' for Plandemic; ``hcq'', ``hydrox'' and ``chloroq'' for hydroxychloroquine; ``mask'' for facemasks; and ``fatality rate'', ``death rate'', ``survival rate'', ``death numbers'', ``covid-19 death'', ``covid death'', ``covid19 death'', ``died from covid'' and ``died of covid'' for covid mortality.
			    
			A random sample of $100$ tweets from each cluster for each topic was selected for coding. If a cluster did not have $100$ tweets on a particular topic, we took all tweets from that cluster on that topic. When possible, each community within a cluster was equally represented in that cluster's random sample, so that a single community did not disproportionately impact the coding results. 
			    
			The coders consisted of four undergraduates from a Midwestern private university. Coders were given a reference sheet (see Supplementary Materials) and a collection of tweet texts with the time the tweet was posted. The time was provided so that coders could cross-reference with the scientific consensus 
about the topic at the time the tweet was posted.  All tweets were presented through an untimed Qualtrics survey. 
			
		\subsection{COVID-19 vaccines and disease severity content}\label{sec:meth-content}
		
			Tweets pertaining to COVID-19 severity and vaccination were identified through a substring search of COVID-19-related tweets (those found to contain any of the the strings ``covid'', ``coronavirus'', ``sars-cov'' or ``pandemic''). Search strings for COVID-19 severity were seeded by reading the COVID-19 tweets and searching for phrases related to the morbidity and mortality of COVID-19, for example those that mention disease incidence or prevalence (e.g. `case spike' or `confirmed cases') as well as death count statistics (e.g. `death count'). For vaccination we used `vaccine', the stem `vaccinat' as well as the names of the vaccine manufacturers whose vaccines received approval from the U.S. Food and Drug Administration. Exact substring lists can be found in the Supplementary Materials.
					
			To identify which topical tweets may be related to misinformation surrounding severity or vaccinations we constructed additional substring lists related to downplaying COVID-19 severity, vaccine hesitancy and vaccine misinformation which were used to filter severity and vaccination tweets respectively.  Complete lists can be found in the Supplementary Materials. 
			
		\subsection{Flagging inter-cluster content spread}\label{sec:meth-sim}
			Inter-cluster content similarity was measured  using a cosine similarity score \cite{manning1999} applied to the trigrams generated by each community's tweets. Specifically, the `documents' used to generate the similarity score were all COVID-19 tweets (Section \ref{sec:meth-coding}) sent by a particular community on a given day. These tweets were cleaned to remove stopwords, urls and mentions of Twitter users, and then trigram frequency vectors were generated using the NLTK package in Python \cite{bird2009}. We define the daily similarity between clusters $A$ and $B$ as the arithmetic mean of the similarity scores between distinct pairs of communities in $A$ and $B$. 
			
			A day was flagged as containing viral content if the similarity between two clusters was anomalously large according to the `burst-score' given in Equation \eqref{eq:hist-sim}. This metric is an adaptation of the burst-score introduced by \cite{mehrotra2013} for inter-cluster similarities.  Specifically, \eqref{eq:hist-sim} measures between similarity on day $t$ in terms of standard deviations from the historic mean. Implicit in this measure is the assumption that the similarity between clusters $A$ and $B$ is stationary in time and does not exhibit a trend.  Validity of this stationarity assumption was assessed with an augmented Dickey-Fuller test \cite{dickey1979}, performed with the \texttt{adfuller} model in the statsmodels \cite{statsmodels} Python package using a `constant only' model with $0$ lag (see Supplementary Materials).

			Latent semantic analysis (LSA) \cite{deerwester1990} was used to identify topical tweets that drove high similarity on days in which at least one pair of clusters was flagged.  LSA was separately applied to the tweets posted by both flagged clusters, where documents corresponded to individual tweets and terms corresponded to the tweet trigrams. The singular document vectors associated with the five largest singular values were used to identify topical tweets from each cluster on the flagged day.  Specifically, a sharp drop in the magnitude of the document vector components was identified and those tweets with component magnitudes above the drop were selected as topical tweets. Topical tweets common to both clusters were removed and the between cluster similarity was recalculated.  Removed tweets were considered to be drivers of the burst on that day if the recalculated burst score was lower than the flagging criterion.

	\section*{Acknowledgements}
	The authors would like to thank Rod Abhari and David King for helpful discussions.  This work was supported by the Office of Research at the Ohio State University. 
	
	\section*{Data Availability}
	Tweet IDs and user IDs corresponding to sentinel nodes that were verified accounts as of July 22, 2020 are available at: \url{https://github.com/joetien/sentinel-node-misinfo}.  
						
\bibliographystyle{ieeetr}
\bibliography{mybib3.bib}


\end{document}